\documentclass[twocolumn,showpacs,preprintnumbers,amsmath,amssymb]{revtex4}

\usepackage{graphicx}% Include figure files
\usepackage{bm}% bold math

\begin{document}

\preprint{Phys. Rev. Lett. (to be published).}

\title{Entanglement of Solid Vortex Matter: A Boomerang Shaped Reduction Forced by Disorder in Interlayer Phase Coherence in Bi$_2$Sr$_2$CaCu$_2$O$_{8+y}$}

\author{T. Kato,$^{1}$ T. Shibauchi,$^{1}$ Y. Matsuda,$^{1,2}$ J.~R. Thompson,$^{3,4}$ and L. Krusin-Elbaum$^5$
}

\affiliation{$^1$Department of Physics, Kyoto University,
Sakyo-ku, Kyoto 606-8502, Japan\\
$^2$Institute for Solid State Physics, University of Tokyo,
Kashiwa, Chiba 277-8581, Japan\\
$^3$Department of Physics and Astronomy, The University of Tennessee,
Knoxville, Tennessee 37996, USA\\
$^4$Materials Science and Technology Division,
Oak Ridge National Laboratory, Oak Ridge, Tennessee 37831, USA\\
$^5$IBM T.~J. Watson Research Center, Yorktown Heights, New York
10598, USA }

\date{Received 2 April 2008; accepted 13 June 2008}

\begin{abstract}
We present evidence for entangled solid vortex matter in a glassy
state in a layered superconductor Bi$_2$Sr$_2$CaCu$_2$O$_{8+y}$
containing randomly splayed linear defects. The interlayer phase
coherence---probed by the Josephson plasma resonance---is enhanced
at high temperatures, reflecting the recoupling of vortex liquid by
the defects. At low temperatures in the vortex solid state, the
interlayer coherence follows a boomerang-shaped reentrant temperature path
with an unusual low field decrease in coherence, indicative of meandering
vortices.
We uncover a distinct temperature scaling between in-plane and
out-of-plane critical currents with opposing dependencies on field
and time, consistent with the theoretically proposed ``splayed-glass''
state.
\end{abstract}

\pacs{74.25.Qt, 74.50.+r, 74.62.Dh, 74.72.Hs}
%74.25.Qt   Vortex lattices, flux pinning, flux creep
%74.50.+r   Tunneling phenomena; point contacts, weak links, Josephson effects (for SQUIDs, see 85.25.Dq; for Josephson devices, see 85.25.Cp; for Josephson junction arrays, see 74.81.Fa)
%74.62.Dh   Effects of crystal defects, doping and substitution
%74.72.Hs   Bi-based cuprates

\maketitle

Entangled states of matter are likely in physical systems where disorder
is ubiquitously present. A widely considered example is that of one
dimensional (1-D) elastic strings subject to random or correlated
defects \cite{Halpin}---a situation encountered in environments as
disparate as directed polymers \cite{Kardar}, magnetic domain walls
\cite{Krusin01}, fire fronts \cite{Maunuksela}, or vortex lines in
superconductors \cite{Review}. The last presents a rich
laboratory for exploring the intricate competition between the
vortex interactions, fluctuations, disorder effects,
and the ability to cut and reconnect---all ultimately
informing the state of coherence and entanglement of lines.

In high transition temperature ($T_{\rm c}$) superconductors, the
competing influences lead to new states of vortex
matter. Bragg glass \cite{Nattermann} with quasi-long-range
translational order can arise in the presence of weak point defects,
while a conceptually different Bose glass state \cite{Nelson}
is promoted by topologically correlated columnar defects
(CDs), where the vortices become unidirectionally localized. Another
distinct ``splayed glass'' state has been predicted \cite{Hwa} in
the presence of directionally distributed CDs; this state 
differs dynamically from other glassy states, with forced
entanglement of vortex lines by the splayed CDs as its chief
characteristic.

The differentiating factor and a key to the nature of these states
reside in the longitudinal vortex correlations.
In layered superconductors, the quasi 2-D pancake vortices
in the layers are connected by Josephson strings to form the lines.
There, vortex correlations
are closely related to the interlayer phase coherence (IPC) $\langle \cos
\phi_{n, n+1} \rangle$, with $\phi_{n, n+1}$ being the
gauge-invariant phase difference between the neighboring
layers $n$ and $n+1$.
IPC can be directly probed by Josephson plasma resonance (JPR)
\cite{Matsuda95,Matsuda97,Shibauchi97,Bulaevskii}, with resonance
frequency $\omega_{\rm p}$ given by
\begin{equation}
\omega_{\rm p}^2(H,T) = \omega_{\rm p}^2(0,T) \langle \cos \phi_{n,
n+1} \rangle = {8\pi^2cs \over \epsilon_0 \Phi_0}J_{\rm c}^c,
\label{omega}
\end{equation}
where $s$ is the interlayer spacing, $\epsilon_0$ the
dielectric constant, $\Phi_0$ the flux quantum, and $J_{\rm c}^c$
the $c$-axis critical current density.

In pristine Bi$_2$Sr$_2$CaCu$_2$O$_{8+y}$ (BSCCO), it has been
demonstrated \cite{Shibauchi,Gaifullin} that the IPC suddenly
decreases with field at the phase transition between the low-field
Bragg glass and the high-field liquid phases, and that within
the liquid phase vortices in each layer are well decoupled
\cite{Koshelev}. In BSCCO with the unidirectional CDs,
known for their strong vortex pinning \cite{Civale,Konczykowski},
an enhanced IPC has been observed 
in the liquid state near $B_\Phi/3$ ($B_\Phi$ is the
matching field where the vortex density coincides with the defect
density) \cite{Sato,Kosugi9799,Tsuchiya,Colson}. Thus, the decoupled
vortex liquid transforms to a recoupled liquid state at higher
fields, where the vortices tend to be confined and forced to align
along the CDs. Forced topological entanglement has been deduced from
transport in the liquid state of YBa$_2$Cu$_3$O$_7$
\cite{Lopez}, where strong interlayer coupling prevents
experimental access to IPC. In more anisotropic BSCCO
\cite{Krusin94} and Hg-based systems \cite{Krusin98}, partial
evidence for entanglement has been claimed from large critical
current enhancements in polycrystalline samples with splayed CDs.
Thus far, however, there is no direct demonstration of entangled
vortex structure via IPC.

Here we report on the IPC in BSCCO single crystals
containing directionally distributed CDs. We
demonstrate that the interlayer coherence in BSCCO with uniformly
{\em splayed} CDs shows an anomalous `boomerang'-like reduction at
low temperatures and low fields. The forced entanglement is
evidenced by the enhanced in-plane critical current density $J_{\rm
c}^{ab}$ with the concurrent drop of $J_{\rm c}^c$---the latter
being a measure of vortex meandering. The found anticorrelation of
the field and time dependence of $J_{\rm c}$'s in the boomerang
region of the phase space is in correspondence with the proposed
framework in the splayed glass state \cite{Hwa}.

%\section{Experimental}

Optimally-doped BSCCO single crystals were prepared by the floating
zone method. Crystals with thicknesses (along the $c$ axis) of $\sim
20~\mu$m were exposed to 0.8~GeV proton beam
in Los Alamos Neutron Science Center (LANSCE). Bi nuclei within the
crystals are fissioned by collisions with swift protons and the
fission fragments create extended randomly oriented damage tracks
with diameter of $\sim 7$~nm and length of $\sim 6~\mu$m
\cite{Krusin94}. In this study, we used two crystals with matching
fields $B_\Phi \approx 0.5$ and 1~T, as estimated from the relation
between areal defect density and the proton fluence \cite{Krusin94}.
JPR experiments were performed by a cavity
perturbation technique at a frequency $\omega/2\pi\approx 28$~GHz
\cite{Shibauchi97,Shibauchi,Sato}, from which we evaluate $J_{\rm c}^{c}$
by Eq.~(\ref{omega}). Standard magnetization measurements
were used to determine the creep
rate $S=-{\rm d}\ln(J_{\rm c}^{ab})/{\rm d}\ln(t)$ as well as
$J_{\rm c}^{ab}$ as a function of field and time \cite{Krusin98}.

%%%%%%%%%%%%%%%%%%%%%%%%%%%%%%%%%%%FIG 1%%%%%%%%%%%%%%%%%%%%
\begin{figure}%[td]
\includegraphics[width=75mm]{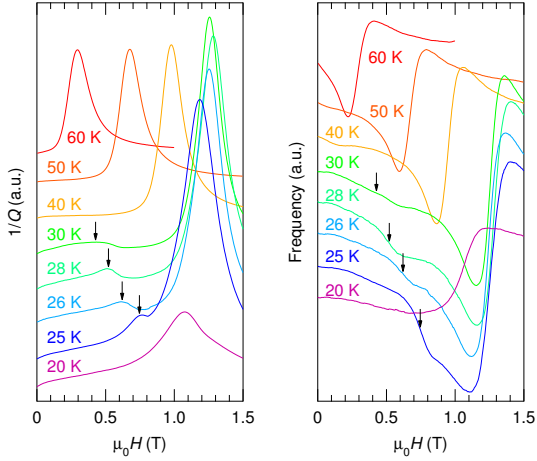}%
\caption{(color online).
Microwave dissipation $1/Q$ (left) and frequency shift
(right) of the 28-GHz Cu cavity including the BSCCO crystal with splayed CDs
($B_\Phi \approx 1$~T) as a function of $H$ applied
parallel to the $c$ axis. Each curve is vertically shifted for clarity.
Arrows mark the second resonance. }
\label{JPR}
\end{figure}
%%%%%%%%%%%%%%%%%%%%%%%%%%%%%%%%%%%FIG 1%%%%%%%%%%%%%%%%%%%%

%\section{Results}
{\it Josephson Plasma Resonance.}--- Figure~\ref{JPR} shows the
field dependence of the inverse quality factor $1/Q$
($\propto$ microwave dissipation) and the
frequency shift for the $B_\Phi \approx 1$~T sample. Magnetic field $H$
is increased after the zero-field-cooling condition. Clear JPR
is observed as a peak in $1/Q$ and a step-like
change in the frequency shift. At high temperatures, we always
observe a single resonance line, in contrast to the case
of BSCCO with parallel CDs along the $c$ axis, where
multiple resonances have been observed near $B_\Phi/3$
\cite{Sato,Kosugi9799,Tsuchiya}.

At low temperatures, in addition to
the main resonance line, we find another resonance (indicated by the
arrows) at a lower field. In this second resonance, the frequency
shift shows a step-like anomaly having an \emph{opposite} direction
from that of the main resonance. This goes to prove that the second
resonance does not originate from some secondary low-$T_{\rm c}$
phase or inhomogeneity, but is an intrinsic feature of this system,
as will be discussed later.

%%%%%%%%%%%%%%%%%%%%%%%%%%%%%%%%%%%FIG 2%%%%%%%%%%%%%%%%%%%%
\begin{figure}%[td]
\includegraphics[width=85mm]{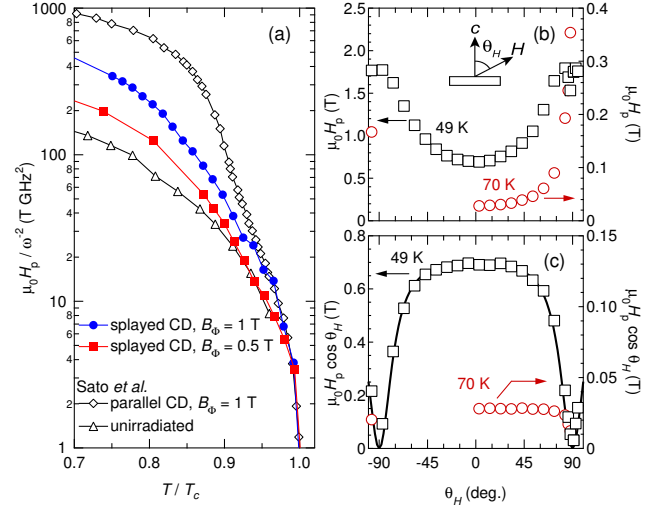}%
\caption{(color online). (a) Resonance field $H_{\rm p}$ normalized
by $\omega^{-2}$ as a function of $T/T_{\rm c}$ for $B_\Phi \approx
1$~T (solid circles) and 0.5~T (solid squares) compared with those
of pristine (open triangles) and heavy-ion irradiated BSCCO with
$B_\Phi=1$~T (open diamonds) \cite{Sato}.
(b) Angle dependence of $H_{\rm p}$
for the high-field state at 49 K (squares) and for the low-field
state at 70 K (circles). $\theta_H$ is the angle between the applied
field and the $c$ axis. (c) $c$-axis component of $H_{\rm p}$ 
vs. $\theta_H$. The solid line is a fit to
Eq.~(\ref{fit}) with $\gamma=400$ and $A=0.47$~deg. } \label{highT}
\end{figure}
%%%%%%%%%%%%%%%%%%%%%%%%%%%%%%%%%%%FIG 2%%%%%%%%%%%%%%%%%%%%

{\it High-Temperature Liquid State.}--- Let us first discuss the
single resonance in the liquid state at high temperatures.
Figure~\ref{highT}(a) shows the temperature dependence of the
resonance field $H_{\rm p}$. To compare this with previous reports
\cite{Sato}, we
normalize the temperature $T$ by $T_{\rm c}$ and $H_{\rm p}$ by
$\omega^{-2}$ \cite{Matsuda95,Shibauchi}.
Near $T_{\rm c}$, our data roughly follow the data for pristine
BSCCO as well as for the one with parallel CDs \cite{Colson}.
However, at low $T$ and high fields a distinct deviation is
observed, indicating that the IPC in the system with randomly splayed CDs
is enhanced relative to the pristine one. A quantitative comparison
with a system containing the parallel CDs with the same $B_\Phi$
shows that the IPC enhancement in our case is considerably smaller.
Moreover, we do not observe multiple resonances in
the liquid state, which are only expected for largely enhanced IPC
with non-monotonic field dependence of $\langle \cos \phi_{n, n+1}
\rangle(H)$ \cite{Kameda}.

The angular field dependence is particularly telling. In BSCCO with
CDs parallel to the $c$ axis, it has been shown that in the
\emph{low-field decoupled} liquid state, the resonance field is
determined mostly by the $c$-axis component over a wide range of
field angles \cite{Kosugi9799}. By contrast, in the \emph{high-field
recoupled} liquid state $H_{\rm p}$ is nearly angle-independent;
indeed, this is a key signature of recoupling.
Our low-field data at 70~K in Fig.~\ref{highT}(b) reproduces the
previous results on the low-field decoupled liquid state; its
$c$-axis component $H_{\rm p}\cos\theta_H$ is almost constant over a
wide angle range. On the other hand, the high-field data at 49~K
shows an odd behavior; neither $H_{\rm p}$ nor $H_{\rm
p}\cos\theta_H$ are angle independent [Figs.~\ref{highT}(b) and
(c)].

%%%%%%%%%%%%%%%%%%%%%%%%%%%%%%%%%%%FIG 3%%%%%%%%%%%%%%%%%%%%
\begin{figure}[t]
\hspace{-5mm}\includegraphics[width=90mm]{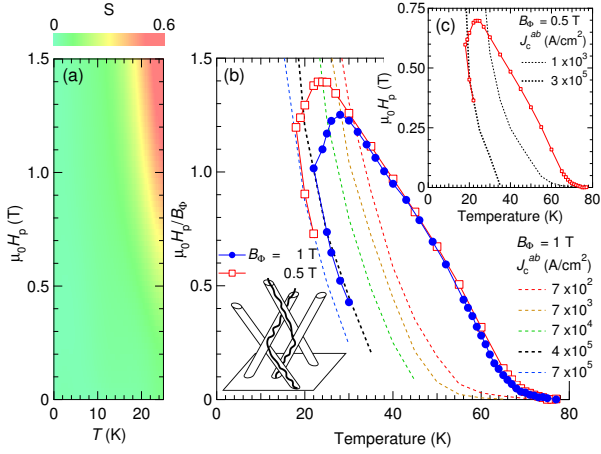}%
\caption{(color online). (a) Contour map of the creep rate $S$ in
$H$-$T$ plane. (b) Temperature dependence of $H_{\rm p}$ normalized
by $B_\Phi$ \cite{resonance}.
Also shown are the contours of in-plane $J_{\rm c}^{ab}$
(dashed lines) for $B_\Phi\approx 1$~T sample.
Sketch of splayed defects and entangled vortices is
also shown. (c) Temperature dependence of $H_{\rm p}$ together with
$J_{\rm c}^{ab}$ contours for $B_\Phi\approx 0.5$~T sample.
} \label{lowT}
\end{figure}
%%%%%%%%%%%%%%%%%%%%%%%%%%%%%%%%%%%FIG 3%%%%%%%%%%%%%%%%%%%%

To analyze this, we apply anisotropic scaling procedure
\cite{Blatter,Krusin98} to the distribution of splayed CDs. The
polar angle $\theta$ from the $c$ axis in anisotropic
superconductors with anisotropy parameter
$\gamma=(m_c/m_{ab})^{1/2}$ can be transformed into $\tilde\theta$
in the scaled isotropic system by $\gamma\tan\tilde\theta =
\tan\theta$. The angle dependence of the trapping probability of
vortices in the scaled isotropic system is described by a Lorentz
function $P(\tilde\theta_H)=1/(1+(\tilde\theta_H/A)^2)$, where the
parameter $A$ related to the accommodation angle is determined
by the pinning strength of the defects \cite{Kameda}.
In our splayed system, the directions of the CDs are
random and the angular dependence of the probability in the
anisotropic system is described by
\begin{equation}
P(\theta_H)=\int_0^{2\pi}\int_0^{\pi/2}
{\sin\theta_{\rm CD} {\rm d}\theta_{\rm CD} {\rm d}\phi_{\rm CD}
\over
1+(\tilde\theta_{H-{\rm CD}}/A)^2}.
\label{fit}
\end{equation}
Here $\theta_{\rm CD}$ is the polar angle between a CD and the $c$
axis, and $\phi_{\rm CD}$ is its angle of longitude with respect to
the field rotation plane. Relative angle between the field and the
defect in the scaled isotropic system is given by
$\tilde\theta_{H-{\rm CD}}=
\cos^{-1} (\sin\tilde\theta_H\sin\tilde\theta_{\rm CD}
\cos\tilde\phi_{\rm CD}
+ \cos\tilde\theta_H\cos\tilde\theta_{\rm CD})$
with $\tilde\theta_H=\tan^{-1}(\tan\theta_H/\gamma)$,
$\tilde\theta_{\rm CD}=\tan^{-1}(\tan\theta_{\rm CD}/\gamma)$, and
$\tilde\phi_{\rm CD}=\phi_{\rm CD}$. As shown in
Fig.~\ref{highT}(c), the $c$-axis component of the resonance field
at 49 K can be fitted to Eq.~(\ref{fit}) with
parameters physically sensible \cite{Kameda}.
Thus, in the high-field liquid state vortices are
largely trapped inside the CDs. We conclude then that here a
recoupling crossover takes place in a field range much smaller than
the matching field $B_\Phi$, much akin to the recoupling induced by
parallel CDs.

%%%%%%%%%%%%%%%%%%%%%%%%%%%%%%%%%%%FIG 4%%%%%%%%%%%%%%%%%%%%
\begin{figure}[t]
\includegraphics[width=90mm]{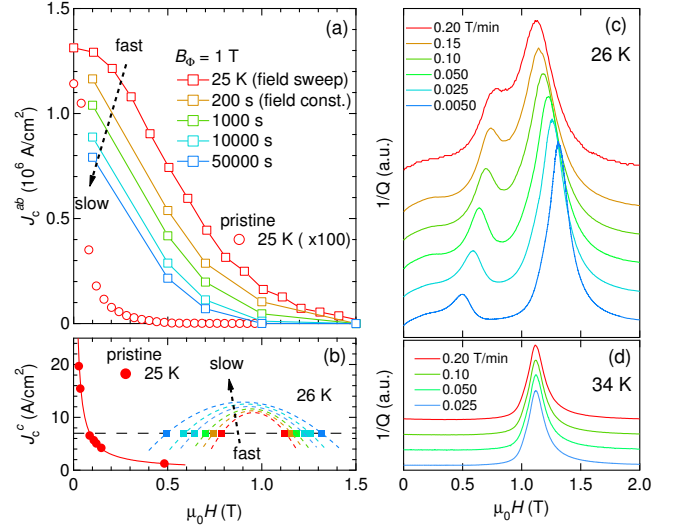}%
\caption{(color online). (a) In-plane $J_{\rm c}^{ab}$ as a function of field
at different times for $B_\Phi\approx 1$~T (squares) and
pristine samples (circles). (b) Field dependence of
$c$-axis $J_{\rm c}^{c}$ for $B_\Phi\approx 1$~T (squares) with different
field sweep rates shows a sharp contrast to the
monotonic $J_{\rm c}^{c}(H)$ in a pristine crystal \cite{Shibauchi} 
with smaller interlayer coupling (circles).
Our measurement frequency corresponds to $J_{\rm c}^{c}\approx
7$~A/cm$^2$. The lines are guides for the eyes.
The JPR fields change with the sweep rate at 26 K (c)
but stay constant at 34 K (d).
The two resonances for each rate in (c)
correspond to the two points in (b) in the same color
\cite{resonance}. In (c) and (d) data are shifted
vertically.
} \label{Jc}
\end{figure}
%%%%%%%%%%%%%%%%%%%%%%%%%%%%%%%%%%%FIG 4%%%%%%%%%%%%%%%%%%%%

{\it Low-Temperature Glass State.}--- In Fig.~\ref{lowT}(b) we plot
the temperature dependence of $H_{\rm p}$ normalized by $B_\Phi$ for
both samples. At high temperatures the two curves coincide,
indicating that the IPC in this system is mostly determined by the
defects. At low temperatures below $\sim 30$~K, $H_{\rm p}(T)$
undergoes a kink and then turns to decrease. This kink is consistent
with previous reports in BSCCO \cite{Matsuda97}, where $H_{\rm
p}(T)$ in the vortex solid state drops almost linearly with
decreasing $T$ in the zero-field-cooling condition. Remarkably,
$H_{\rm p}(T)$ in our system has an anomalous \emph{reentrant}
behavior---it is this boomerang-like trajectory that gives double
resonance lines in single field sweep runs at temperatures below
$\sim 30$~K \cite{resonance}.

Since we use a fixed microwave frequency and $\omega_{\rm p}(0,T)$
in Eq.~(\ref{omega}) is only weakly temperature dependent at low
temperatures \cite{Shibauchi}, the resonance field curve $H_{\rm
p}(T)$ provides a rough contour of IPC at low temperatures. Thus,
the observation of the reentrant behavior at low temperatures in
this system demonstrates that the field dependence of $\langle \cos
\phi_{n, n+1} \rangle$ at a fixed temperature is non-monotonic [see
also $J_{\rm c}^c(H)$ in Fig.~\ref{Jc}(b)]: at low fields it
increases with field and at high fields it decreases. The observed
opposite steps at the two resonances in the frequency shift in
Fig.~\ref{JPR} are in accord with this result.

More importantly, $H_{\rm p}(T)$ at low fields follows a contour of
in-plane $J_{\rm c}^{ab}\approx 3$-$4 \times10^5$~A/cm$^2$
[Figs.~\ref{lowT}(b) and (c)] as well as of the creep rate $S$
[Fig.~\ref{lowT}(a)]. This newly found scaling can be naturally
understood when the meandering vortex structure (witnessed by the
low $H_{\rm p}$) facilitates its pinning and reduces the vortex
dynamics.

In the proposed ``splayed glass'' phase \cite{Hwa},
vortices entangle by confining their finite segments to different columnar
paths [see sketch in Fig.~\ref{lowT}(b)].
The dynamics is determined by the nucleation and the motion of kinks,
which are impeded from sliding along the CDs at the crossing sites when the
CDs are roughly localization length apart \cite{Hwa,Schuster}.
In the uniformly splayed CD landscape, random exchanges of vortex lines
with numerous kinks will show up as a reduction of IPC,
which is proportional to $J_{\rm c}^{c}$ [see Eq.~(\ref{omega})].

In Figs.~\ref{Jc}(a) and (b), we compare $J_{\rm c}^{ab}(H)$ and
$J_{\rm c}^{c}(H)$ with those in the pristine samples.
From the boomerang $H_{\rm p}(T)$ in Fig.~\ref{lowT}, at a
fixed temperature we expect a non-monotonic $J_{\rm c}^{c}(H)$.
Indeed, in sharp contrast with the pristine samples, at low fields
$J_{\rm c}^{c}$ decreases with decreasing $H$ [dashed lines in
Fig.~\ref{Jc}(b)]. Concurrently, the in-plane
$J_{\rm c}^{ab}$ rapidly increases, fast exceeding that of
pristine one. We remark that this $J_{\rm c}$ anti-correlation is
counterintuitive in a coupled pancake system and unique to the
splayed glass phase \cite{note}. In pristine BSCCO, the reduction of
longitudinal correlation due to decoupling is a consequence of
in-plane correlations becoming dominant at higher fields. We also
note that in the Bose glass state discussed in the system
with parallel CDs such an anti-correlation between $J_{\rm
c}^{c}$ and $J_{\rm c}^{ab}$ is not expected. Thus the low-field
$J_{\rm c}^{c}$ reduction discovered in our system is anomalous.

Also striking is that the two resonance lines
split apart [Fig.~\ref{Jc}(c)]---a slower field-sweep rate
gives rise to a larger $J_{\rm c}^{c}$ [Fig.~\ref{Jc}(b)].
We emphasize that the time relaxation of $J_{\rm c}^{ab}$ shows an opposite
trend [Fig.~\ref{Jc}(a)].
All these findings are characteristics of splayed glass.

Thus, through our experiments we demonstrate unusual `line'
behaviors in a system with randomly oriented linear defects. The
non-monotonic field dependence of $\langle \cos \phi_{n, n+1}
\rangle$, the reentrant ``boomerang'' path of the JPR field,
the scaling of this path in the $H$-$T$ phase space
with the contours of iso-$J_{\rm c}^{ab}$, all support a strong
low-field suppression of interlayer phase coherence. While a full
theory of $\langle \cos \phi_{n, n+1} \rangle$ in a system of random
distributions of CDs that includes point disorder is yet
to be formulated, our findings in vortex structure and dynamics
support the presence of a distinct ``splayed glass'' entangled state
of matter.

%\section*{Acknowledgments}
We thank V.~M. Vinokur, A. Koshelev, C.~J. van der Beek,
M. Konczykowski for discussion,
and J. Ullmann for assistance at LANSCE.
ORNL research was sponsored by Div. of Materials Sciences
and Engineering, U.S. DOE.


\begin{references}

%\item[*] Email address: shibauchi@scphys.kyoto-u.ac.jp

\bibitem{Halpin} T. Halpin-Healy and Y.-C. Zhang,
Phys. Rep. {\bf 254}, 215 (1995).

\bibitem{Kardar} M. Kardar and Y.-C. Zhang,
Phys. Rev. Lett. {\bf 58}, 2087 (1987).

\bibitem{Krusin01} L. Krusin-Elbaum, T. Shibauchi, B. Argyle, L. Gignac,
and D. Weller, Nature {\bf 410}, 444 (2001);
T. Shibauchi {\it et al.},
Phys. Rev. Lett. {\bf 87}, 267201 (2001).

\bibitem{Maunuksela} J. Maunuksela {\it et al.},
Phys. Rev. Lett. {\bf 79}, 1515 (1997).

\bibitem{Review} G. Blatter, M.~V. Feigel'man, V.~B. Geshkenbein,
A.~I. Larkin, and V.~M. Vinokur,
Rev. Mod. Phys. {\bf 66}, 1125 (1994).

\bibitem{Nattermann} T. Nattermann and S. Scheidl,
Adv. Phys. {\bf 49}, 607 (2000);
T. Giamarchi and P. Le~Doussal,
Phys. Rev. B {\bf 52}, 1242 (1995).

\bibitem{Nelson} D.~R. Nelson and V.~M. Vinokur,
Phys. Rev. B {\bf 48}, 13060 (1993).

\bibitem{Hwa} T. Hwa, P. Le Doussal, D.~R. Nelson, and V.~M. Vinokur,
Phys. Rev. Lett. {\bf 71}, 3545 (1993).

\bibitem{Matsuda95} Y. Matsuda, M.~B. Gaifullin, K. Kumagai,
K. Kadowaki, and T. Mochiku,
Phys. Rev. Lett. {\bf 75}, 4512 (1995).

\bibitem{Matsuda97} Y. Matsuda, M.~B. Gaifullin, K.~I. Kumagai,
M. Kosugi, and K. Hirata,
Phys. Rev. Lett. {\bf 78}, 1972 (1997).

\bibitem{Shibauchi97} T. Shibauchi {\it et al.},
%M. Sato, A. Mashio, T. Tamegai, H. Mori, S. Tajima, S. Tanaka,
Phys. Rev. B {\bf 55}, R11977 (1997);
T. Shibauchi, M. Sato, S. Ooi, and T. Tamegai,
Phys. Rev. B {\bf 57}, R5622 (1998).

\bibitem{Bulaevskii} L.~N. Bulaevskii, M.~P. Maley, and M. Tachiki,
Phys. Rev. Lett. {\bf 74}, 801 (1995).

\bibitem{Shibauchi} T. Shibauchi {\it et al.},
%T. Nakano, M. Sato, T. Kisu, N. Kameda, N. Okuda, S. Ooi, T. Tamegai,
Phys. Rev. Lett. {\bf 83}, 1010 (1999).

\bibitem{Gaifullin} M.~B. Gaifullin, Y. Matsuda, N. Chikumoto,
J. Shimoyama, and K. Kishio,
Phys. Rev. Lett. {\bf 84}, 2945 (2000).

\bibitem{Koshelev} A.~E. Koshelev,
Phys. Rev. Lett. {\bf 77}, 3901 (1996).

\bibitem{Civale} L. Civale {\it et al.},
%A.~D. Marwick, T.~K. Worthington, M.~A. Kirk, J.~R. Thompson,
%L. Krusin-Elbaum, Y. Sun, J.~R. Clem, F. Holtzberg,
Phys. Rev. Lett. {\bf 67}, 648 (1991).

\bibitem{Konczykowski} M. Konczykowski {\it et al.},
%F. Rullier-Albenque, E.~R. Yacoby, A. Shaulov, Y. Yeshurun, P. Lejay,
Phys. Rev. B {\bf 44}, 7167 (1991).

\bibitem{Sato} M. Sato, T. Shibauchi, S. Ooi, T. Tamegai, and
M. Konczykowski,
Phys. Rev. Lett. {\bf 79}, 3759 (1997).

\bibitem{Krusin98} L. Krusin-Elbaum {\it et al.},
%G. Blatter, J.~R. Thompson, D.~K. Petrov, R. Wheeler, J. Ullmann, C. W. Chu,
Phys. Rev. Lett. {\bf 81}, 3948 (1998).

\bibitem{Kosugi9799} M. Kosugi {\it et al.},
%Y. Matsuda, M.~B. Gaifullin, L.~N. Bulaevskii, N. Chikumoto,
%M. Konczykowski, J. Shimoyama, K. Kishio, K. Hirata, K. Kumagai,
Phys. Rev. Lett. {\bf 79}, 3763 (1997); %\bibitem{Kosugi99}
M. Kosugi {\it et al.},
%Y. Matsuda, M.~B. Gaifullin, L.~N. Bulaevskii, N. Chikumoto,
%M. Konczykowski, J. Shimoyama, K. Kishio, K. Hirata,
Phys. Rev. B {\bf 59}, 8970 (1999).

\bibitem{Tsuchiya}
T. Hanaguri, Y. Tsuchiya, S. Sakamoto, A. Maeda,
and D.~G. Steel,
Phys. Rev. Lett. {\bf 78}, 3177 (1997):
Y. Tsuchiya {\it et al.},
%T. Hanaguri, H. Yasuda, A. Maeda,
%M. Sasase, K. Hojou, D.~G. Steel, J.~U. Lee, D.~J. Hofman,
Phys. Rev. B {\bf 59}, 11568 (1999).

\bibitem{Colson} S. Colson {\it et al.},
%C.~J. van der Beek, M. Konczykowski, M.~B. Gaifullin, Y. Matsuda,
%P. Gierlowski, M. Li, P.~H. Kes,
Phys. Rev. B {\bf 69}, 180510(R) (2004).

\bibitem{Lopez} D. L\'opez {\it et al.},
Phys. Rev. Lett. {\bf 79}, 4258 (1997).

\bibitem{Krusin94} L. Krusin-Elbaum {\it et al.},
%J.~R. Thompson, R. Wheeler, A.~D. Marwick, C. Li, S. Patel,
%D.~T. Shaw, P. Lisowski, J. Ullmann,
Appl. Phys. Lett. {\bf 64}, 3331 (1994).

\bibitem{Kameda} N. Kameda {\it et al.},
%T. Shibauchi, M. Tokunaga, S. Ooi, T. Tamegai, M. Konczykowski,
Phys. Rev. B {\bf 72}, 064501 (2005).

\bibitem{Blatter} G. Blatter, V.~B. Geshkenbein, and A. I. Larkin,
Phys. Rev. Lett. {\bf 68}, 875 (1992).

\bibitem{resonance}
At zero field, the plasma frequency is expected to be large
($J_{\rm c}^c > 7$~A/cm$^2$), but we do not observe
the third resonance line at low fields.
This is likely due to inhomogeneous field
distribution arising from the large $J_{\rm c}^{ab}$.

\bibitem{Schuster} T. Schuster {\it et al.},
%H. Kuhn, M.~V. Indenbom, G. Kreiselmeyer, M. Leghissa, S. Klaum\"unzer,
Phys. Rev. B {\bf 53}, 2257 (1996).

\bibitem{note}
We note the important role of points defects (most
effective at low $T$) \cite{Matsuda97,Tsuchiya} in reducing the
motion of kinks as well as in reducing the interlayer coherence.

\end{references}
\end{document}